\documentclass[aps, prl, showpacs, twocolumn, amsfonts, amsmath, amssymb, superscriptaddress, floatfix]{revtex4}

\usepackage{graphicx}
\usepackage{dcolumn}
\usepackage{bm}
\usepackage{epsfig}
\usepackage{hyperref}

\begin{document}

\title{Controlled Dicke Subradiance from a Large Cloud of Two-Level Systems}

\author{Tom Bienaim\'e}
\affiliation{Universit\'e de Nice Sophia Antipolis, CNRS, Institut Non-Lin\'eaire de Nice, UMR 7335, F-06560 Valbonne, France}

\author{Nicola Piovella}
\affiliation{Dipartimento di Fisica, Universit\`a Degli Studi di Milano, Via Celoria 16, I-20133 Milano, Italy}

\author{Robin Kaiser}
\affiliation{Universit\'e de Nice Sophia Antipolis, CNRS, Institut Non-Lin\'eaire de Nice, UMR 7335, F-06560 Valbonne, France}

\date{\today}

\begin{abstract}
Dicke superradiance has been observed in many systems and is based on constructive interferences between many scattered waves. The counterpart of this enhanced dynamics, subradiance, is a destructive interference effect leading to the partial trapping of light in the system. In contrast to the robust superradiance, subradiant states are fragile and spurious decoherence phenomena hitherto obstructed the observation of such metastable states. We show that a dilute cloud of cold atoms is an ideal system to look for subradiance  in free space and study various mechanisms to control this subradiance.
\end{abstract}


\maketitle

Interferences between scattered waves by many particles can give rise to a large variety of phenomena, including collective effects such as Bragg or Mie scattering, well known in the context of classical optics \cite{Hulst57}. More intriguing situations can arise in the mesoscopic regime where interferences are at the origin of coherent backscattering and Anderson localization \cite{Akkermans}. Interest in trapping waves in disordered media has spurred the efforts in multiple scattering of light, as photons seem to be an ideal candidate for non-interacting waves. Several mechanisms allowing to increase the time an optical excitation can stay in a system are known to exist and are based on different physical phenomena \cite{Kaiser09, Havey10}. Multiple scattering of light or radiation trapping for instance allows for increased photon trapping in the absence of interferences \cite{Labeyrie06}. Anderson localization is another possibility where interferences form localized states with exponentially decreasing coupling to the environment and a dense sample fulfilling the Ioffe-Regel criterion \cite{Ioffe60} is assumed to be required.
This letter addresses another fundamental mechanism leading to `long-lived' modes of excitation: the study of subradiance in dilute clouds of cold atoms i.e. the trapping of light by destructive interferences in the single scattering regime. This mechanism is based on the pioneering work by Dicke who studied enhanced decay rates in small and large samples \cite{Dicke1954}. Typically Dicke states are considered for an assembly of $N$ two-level systems, realized e.g. by atoms \cite{Gross82} or quantum dots \cite{Lodahl04}. Different regimes can be studied, using either an initially fully inverted system with $N$ photons stored by the $N$ atoms or an initial state in which the whole system shares a single excitation \cite{AGK2008, Scully_Science}. These systems have attracted an increasing attention in the context of quantum information science \cite{Brandes05, Sanders07, Molmer08}, where the accessible Hilbert space can be restricted to single excitations using e.g. the Rydberg blockade \cite{Gould04,Weidemuller04,Urban09,Gaetan09}.

In this letter, we will show how it is possible to understand and control the coupling of light into metastable subradiant states, illustrating that large dilute clouds of cold atoms are an ideal system to observe for the first time long photon storage in a system of $N$ atoms in free space. Subradiance for two ions has been observed in the past \cite{Brewer96} and a reduced decay rate into one radiation mode has been achieved for $N$ atoms \cite{Pillet85}. However, it has not yet been possible to control and suppress the decay into all vacuum modes for $N$ atoms in free space extending thus the lifetime of the excitation to many times the natural lifetime of a single atom. Starting from the Ansatz used in previous work by several authors \cite{Scully_Science, Eberly06, Glauber08, Bienaime10, Courteille10}, we show that the exponential kernel of the dipole-dipole coupling yields an important fraction of atoms to be coupled into subradiant modes. Following \cite{Woggon05}, we study different inhomogeneous broadening mechanisms, which allow to go beyond the fundamental Fano coupling by controlled coupling between super- and subradiant states using Doppler broadening and inhomogeneous light shifts.
In this letter, we consider three different experimental parameters to control subradiance: the optical thickness of the cloud, the driving laser intensity, and the temperature of the cloud. Moreover, we show how to distinguish subradiance from multiple scattering of light by tuning the driving laser frequency.

We consider a Gaussian cloud, with root mean square size $\sigma$, of $N$ two-level atoms (positions $\mathbf r_i$, transition wavelength $\lambda=2\pi/k$, excited state lifetime $1/\Gamma$), excited by an incident laser (Rabi frequency $\Omega_0$, detuning $\Delta_0$, wavevector $\mathbf{k}_0$). We define the optical thickness $b(\Delta_0)=b_0/\left[1+4(\Delta_0/\Gamma)^2 \right]$ with $b_0 =3N/(k\sigma)^2$ its resonant value. Restricting the atomic Hilbert space to the subspace spanned by the ground state of the atoms $|G \rangle \equiv |g \cdots g \rangle$ and the single excited states $|i\rangle \equiv |g \cdots e_i \cdots g \rangle$ and tracing over the photon degrees of freedom, one obtain an effective Hamiltonian describing the time evolution of the atomic wavefunction $|\psi\rangle = \alpha |G\rangle + \sum_i \beta_i |i\rangle$. The effective Hamiltonian using standard approximations \cite{Courteille10,Bienaime11} can then be written as:
\begin{eqnarray}
H_{\text{eff}} &=& \frac{\hbar \Omega_0}{2} \sum_i \left[ e^{i \Delta_0 t - i \mathbf k_0 \cdot \mathbf r_i} S^i_- +  e^{- i \Delta_0 t + i \mathbf k_0 \cdot \mathbf r_i} S^i_+ \right] \nonumber \\
& & - \frac{i \hbar \Gamma}{2}  \sum_i S_+^i S_-^i - \frac{\hbar \Gamma}{2} \sum_{i} \sum_{j \neq i}  V_{ij} S_+^i S_-^j, \label{Heff}
\end{eqnarray}
where the first term describes the coupling to the laser field, the second accounts for the finite lifetime of the excited states, the third one describes the dipole-dipole interactions, with $V_{ij} =  \frac{\exp i k |\mathbf r_i - \mathbf r_j|}{k |\mathbf r_i - \mathbf r_j|}$, and $S_\pm^i$, $S_z^i$ are the 
usual pseudo-spin operators for the kets $|g_i\rangle$ and $|e_i\rangle$. The effective Hamiltonian (\ref{Heff}) describes dipole-dipole couplings in the scalar light approximation, where near field and polarization effects are neglected since we are considering dilute clouds, $N (\lambda / \sigma)^3 \ll 1$.
At low intensity (where the single excitation approximation is valid), $\alpha \simeq 1$ and the previous model describes a system of $N$ classical dipoles driven by an incident electric field as expected by linear optics \cite{Svidzinsky10}. We use these classical equations to study single photon subradiance. The driven steady state solution $|\psi\rangle \simeq |G\rangle + \epsilon |TD\rangle$ bears the phenomenon of single photon superradiance with a `timed Dicke' state $|TD\rangle=\frac{1}{\sqrt{N}}\sum_{i} e^{i \mathbf k_0 \mathbf r_i} |i\rangle$ and an amplitude $\epsilon \simeq \sqrt{N} \Omega_0/(2\Delta_0+i(1+b_0/12)\Gamma)$ \cite{Courteille10}.
This was e.g. exploited to explain the measured cooperative radiation pressure force in dilute clouds \cite{Bienaime10}.

In order to study enhanced storage time based on Dicke subradiance, we will investigate the coupling between `short-lived' superradiant states, such as $|TD\rangle$, and the other states $|\varphi\rangle$ of the single excitation Hilbert subspace.
Let us first consider the minimal coupling, based on radiative dipole-dipole coupling.
As the effective Hamiltonian Eq. (\ref{Heff}) is non-Hermitian, its eigenstates are non orthogonal ($\langle\varphi_{\text{super}}|\varphi_{\text{sub}}\rangle\neq 0$) and subradiant states thus have common features with auto-ionizing states or Fano resonances \cite{Fano61}.
This Fano-type coupling between `long-lived' subradiant states and `short-lived' superradiant states leads to additional decay channels in addition to direct decay of the subradiant states to the ground state. This situation is reminiscent of the Hanle effect \cite{Kaiser91} where a competition of direct decay and transverse coupling can lead to surprisingly narrow resonances.
For instance, the steady state solution of optical Bloch equations for a system consisting of the ground state $|G\rangle$, one superradiant $|\varphi_{\text{super}}\rangle$ and a single subradiant state $|\varphi_{\text{sub}}\rangle$ reveals that in the absence of Fano coupling the steady state solution of the subradiant state is zero. Neglecting the direct decay of the subradiant state to the ground state on the other hand, one finds that for resonant excitation and small coupling between the excited states, after a long transient time all atoms are pumped into the subradiant state $|\varphi_{\text{sub}}\rangle$. The optical Bloch equations of such a simplified three-level model show that, by increasing in a controlled manner the coupling between super- and subradiant states, it should be possible to efficiently store populations into `long-lived' subradiant modes.


\begin{figure}[t]
\centerline{{\includegraphics[height=5cm]{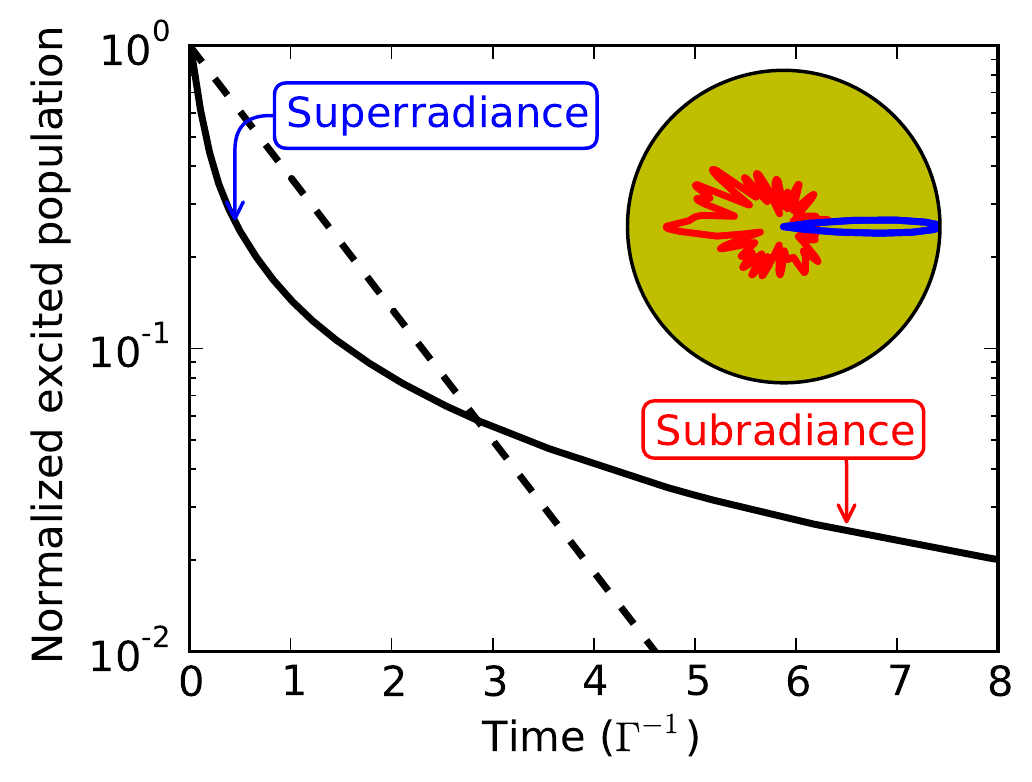}}}
\caption{(color online). Time evolution of the normalized excited state population (black solid curve) after switching off the laser for $N=2000$ atoms, $k \sigma = 10$, $\Delta_0 = 10 \, \Gamma$ (the laser was on before during $50 \, \Gamma^{-1}$ to let the system reach the steady state). At first, the population decreases faster than the single atom decay (black dashed line) and then slower. We respectively identify these phenomena as super- and subradiance. The inset shows the emission diagrams of the superradiant `timed Dicke' state $|TD\rangle$ (blue) and subradiant modes (red) averaged over $8$ realizations (rescaled to allow convenient comparison).}
\label{Fig1}
\end{figure}

In Fig. \ref{Fig1}, we show the normalized excited state population $\propto \sum_i |\beta_i|^2$ as computed from numerical solution of the effective Hamiltonian Eq. (\ref{Heff}) in the linear regime. As precise initial conditions play a crucial role in the subsequent fast and slow decay \cite{Svidzinsky10}, we start initially with all atoms in the ground state $|G\rangle$ and keep the coherent laser drive for $50 \, \Gamma^{-1}$ before switching it off, realizing thus experimentally accessible conditions. The fast initial decay of the superradiant state $\Gamma_{\text{super}}=(1+b_0/12) \, \Gamma$ is clearly seen. Moreover, after this initial fast decay, subradiance manifests itself in a slowly decaying excited population with a rate well below the single atom decay rate. At first, the subradiant decay is not purely exponential since several modes decay simultaneously. For longer times, it then ends up with a pure exponential decay (referred as subradiant decay in the following) when only one `long-lived' mode dominates. The emission diagram of the superradiant `timed Dicke' state $|TD\rangle$ is clearly forward directed (see Fig. \ref{Fig1}), a phenomenon reminiscent of Mie scattering. On the other hand, subradiant modes show isotropic diagrams. They do not possess the symmetry of the laser excitation since they are not directly coupled to it, a feature which can be exploited in the experimental detection of subradiance.

\begin{figure}[t]
\centerline{{\includegraphics[height=5cm]{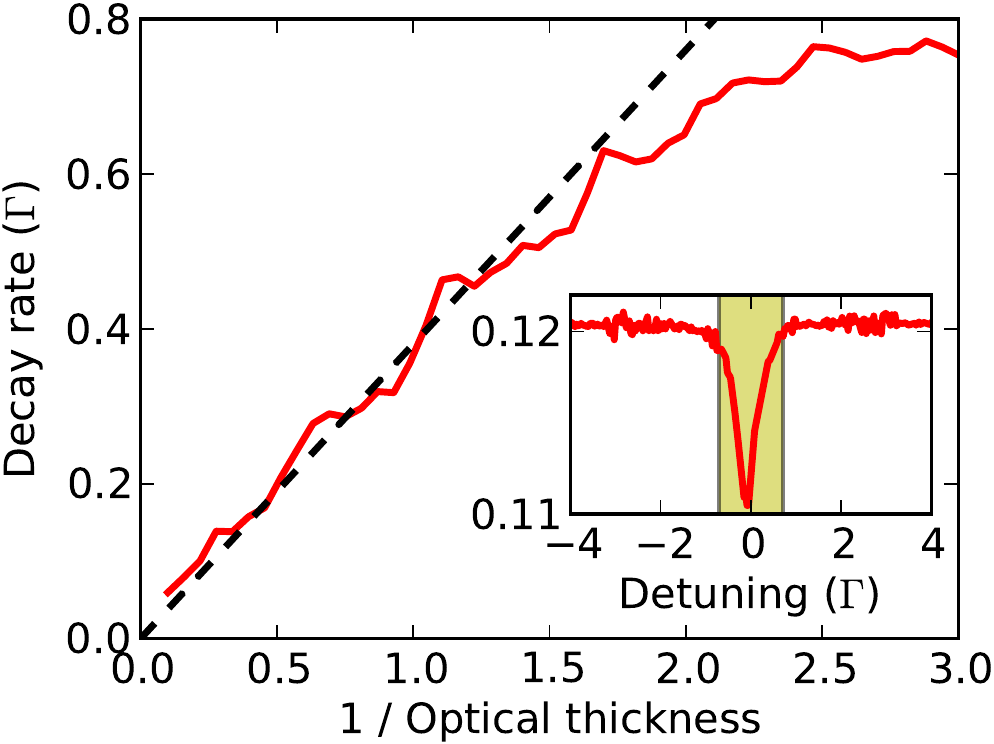}}}
\caption{(color online). Subradiant decay rate as a function of $1/b_0$ (we kept $N=400$ constant, $\Delta_0 = 10 \, \Gamma$ and varied $k \sigma$). At large optical thickness it scales as $\propto \Gamma/b_0$ (black dashed line) and saturates to $\Gamma$ for dilute clouds. The insert shows the subradiant decay rate as a function of detuning for $b_0 = 3$ ($N= 400$, $k \sigma = 20$). The shaded area corresponds to the multiple scattering region $b(\Delta_0) > 1$.}
\label{Fig2}
\end{figure}

In Fig. \ref{Fig2}, the subradiant decay rates (due to Fano coupling) are plotted as a function of the inverse on-resonance optical thickness of the system. The difference to multiple scattering of light which can also yield long photon trapping times, can be understood by looking at the decay rates for different excitation frequencies. As shown in the inset of Fig. \ref{Fig2}, close to resonance we observe reduced decay rates, which we associate to the large optical thickness for resonant photons \cite{Labeyrie06}. For larger detunings however, the decay rate becomes independent of the excitation frequency, consistent with the subradiant nature of states weakly excited off-resonance.

The fundamental Fano coupling between sub- and superradiant states in the Dicke basis can be understood from the diagonal terms in the bare bases $|i\rangle$.
Indeed, the local field at i-th atom is the sum of the external field $E_0$ and the field scattered by all the other dipoles $j$ at the location of the atom $i$:
\begin{equation}
E_{\text{tot}}(\mathbf r_i)=E_0(\mathbf r_i)+\sum_{j \not = i}E_j(\mathbf r_i)\simeq E_0(\mathbf r_i)(1+\varepsilon_i e^{i\varphi_i}).
\label{EqSpeckle}
\end{equation}
We describe the scattered field by a small local speckle field with a random amplitude scaling as $\varepsilon_i \propto \sqrt{b(\Delta_0)} \ll 1$ and a random phase $\varphi_i$.
The amplitudes $\beta_i$ depend on the local field and differ from the driven timed Dicke amplitudes.
The dipole $i$ is thus driven with a random field corresponding to an inhomogeneous broadening mechanism \cite{Woggon05}, where we now have inhomogeneity in amplitude and phase.
An estimation of the expected lifetimes of the `long-lived' subradiant modes is also an important issue.  We have checked numerically they scale as
\begin{equation}
\Gamma_{\text{sub}}\sim \frac{1}{b_0},
\label{Gamma_sub}
\end{equation}
for $b_0 > 1$ (see Fig. \ref{Fig2}).
This scaling is close to what can be obtained assuming that in the limit of large detuning $\Delta_0$, where multiple scattering inside the sample can be neglected, the escape rate of the excitation from the sample is well approximated by the inverse of the `long-lived' mode lifetime, which scale as $1/b_0$ \cite{Louis11}.
This can also be compared to scaling laws obtained from quantum chaotic scattering theory \cite{Sommers99}: for large $b_0$ (corresponding to a large number of atoms $N$ compared to the number of outgoing modes $M\sim (k\sigma)^2$) the minimum width of the resonance is expected to scale as $\Gamma_{\text{sub}}\sim M/N \sim 1/b_0$.


We now turn to the possibility of controlled coupling between super- and subradiant states, opened by cold atoms. In atomic physics many different inhomogeneous broadening mechanisms are known. We will focus on two such mechanisms, well adapted to become a control knob to steer excitations into the subradiant state.
\begin{figure}[t]
\centerline{{\includegraphics[height=5cm]{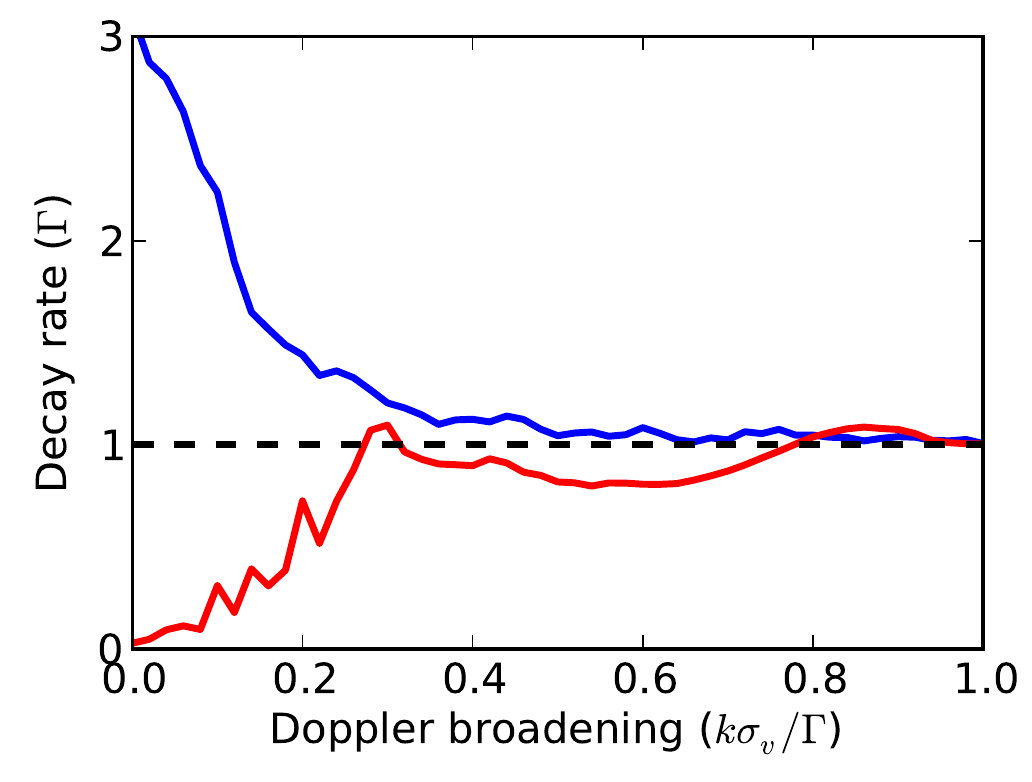}}}
\caption{(color online). Super- (blue curve) and subradiant (red curve) decay rates as a function of Doppler broadening for $N=200$ atoms, $k \sigma=5$ and a laser detuning of $\Delta_0 =10 \, \Gamma$. Initially the laser is driving the atoms for $50 \, \Gamma^{-1}$ and the subradiant decay rate is evaluated $50 \, \Gamma^{-1}$ after the laser is switched off.}
\label{Fig3}
\end{figure}
Let us thus consider the impact of residual motion of the atoms. The effective Hamiltonian (\ref{Heff}) can be extended to include Doppler shifts and time dependent positions of the atoms. The coupled equations for the dipole amplitudes $\beta_i$ read in the linear regime
\begin{align}\label{Doppler}
    \dot{\beta}_i= &\left[-\frac{\Gamma}{2} + i(\Delta_0-\mathbf{k}_0\cdot \mathbf{v}_i) \right] \beta_i - \tfrac{i \Omega_0}{2} +\frac{ i\Gamma}{2}\sum_{j\neq i}
    V_{ij}
    \beta_j,
\end{align}
where $V_{ij}(t)=\frac{e^{ik|\mathbf{r}_i-\mathbf{r}_j+(\mathbf{v}_i-\mathbf{v}_j)t|}}
    {k|\mathbf{r}_i-\mathbf{r}_j+(\mathbf{v}_i-\mathbf{v}_j)t|}
    e^{-i\mathbf{k}_0\cdot[\mathbf{r}_i-\mathbf{r}_j+(\mathbf{v}_i-\mathbf{v}_j)t]}$.
Solving these equations for increasing temperature, we notice that the fast superradiant and the slow subradiant decay rates draw closer to the single atom decay rate $\Gamma$ (see Fig. \ref{Fig3}). This result shows that the fragile subradiant modes are quickly destroyed even by moderate atomic motion. Cold atoms seem a well adapted system allowing to tune the atomic motion from being negligible to becoming dominant. We also checked that the dominant term in the reduced super and subradiance stems from the position dependent dipole-dipole coupling $V_{ij}(t)$ term rather than from the random detuning term in Eq. (\ref{Doppler}). This dependence on the atomic motion explains why subradiance of $N$ atoms has not been observed in hot atomic vapors, despite the efforts in this field in the 1970s \cite{Gross82}.

\begin{figure}[t]
\centerline{{\includegraphics[height=3.6cm]{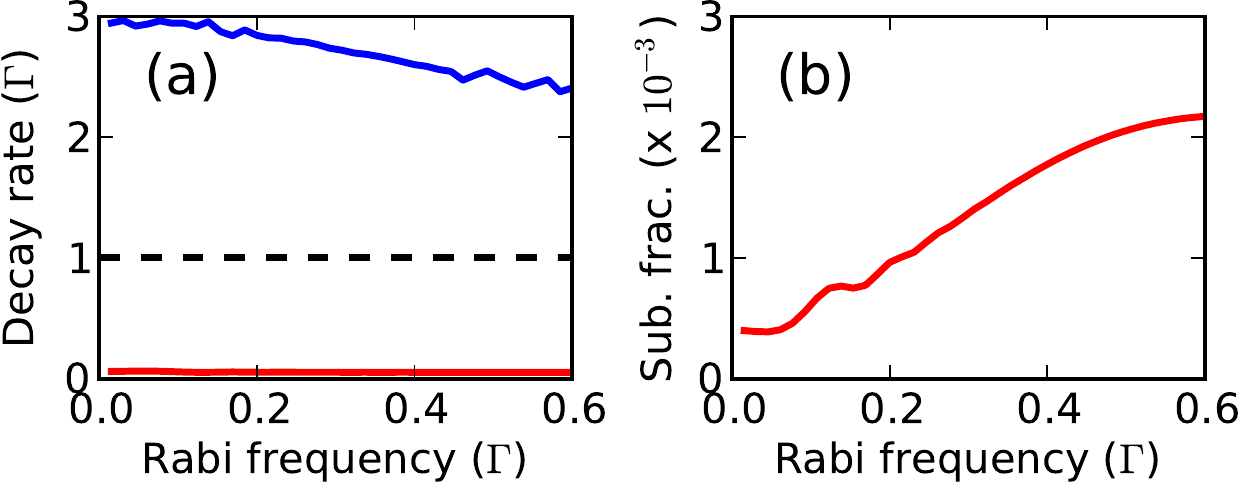}}}
\caption{(color online). (a) Super- (blue curve) and subradiant (red curve) decay rates as a function of laser intensity for $N=200$ atoms, $k\sigma=5$ ($b_0=24$) and a laser detuning of $\Delta_0 =10 \, \Gamma$. (b) Subradiant fraction for the same parameters. The laser is switched  on during $50 \, \Gamma^{-1}$ and the subradiant decay rate and subradiant fraction (i.e. the remaining excited state population) are computed $50 \, \Gamma^{-1}$ after switching off the laser.}
\label{Fig4}
\end{figure}

Exploiting the possibility of controlled coupling via inhomogeneous broadening even further, we studied the role of larger laser intensities on super- and subradiance. As the effective Hamiltonian approach is only valid to first order in the Rabi field coupling, it can only treat the linear optics regime. We therefore used a master equation approach, where the evolution of the atomic density operator $\rho$ in the electric-dipole, rotating-wave, and Born-Markov approximations is given in the interaction picture by \cite{Agarwal74,Carmichael93}
\begin{eqnarray}
\dot \rho &=& \frac{1}{i \hbar} \left( H_{\text{eff}} \rho - \rho H_{\text{eff}}^\dagger \right) + \sum_{i} \sum_{j} \gamma_{ij} S_-^j \rho S_+^i \label{master}
\end{eqnarray}
where in the scalar light limit $\gamma_{ij} = \Gamma \frac{\sin k |\mathbf r_i - \mathbf r_j|}{k |\mathbf r_i - \mathbf r_j|}$. Projecting Eq. (\ref{master}) on the different Fock states, we obtain a set of coupled equations for the density matrix elements $\rho_{G|G} \equiv \langle G | \rho | G \rangle $, $\rho_{i|G} \equiv \langle i | \rho | G \rangle $ and $\rho_{i|j} \equiv \langle i | \rho | j \rangle $. This approach is still restricted to the same Hilbert subspace with at most one excitation and is thus limited to moderate laser intensities or to situations where multiple excitations are suppressed as for instance in the case of Rydberg blockade. However, it does not require the ground state population to remain unaffected and can take into account the light shifts of the states. We have checked analytically and numerically that at first order in Rabi frequency of the laser coupling, the master equation is equivalent to the equation used in the effective Hamiltonian approach Eq. (\ref{Heff}) in the linear regime $\alpha \simeq 1$ (with correspondence $\rho_{G|G} \leftrightarrow 1$, $\rho_{i|G} \leftrightarrow \beta_i$).
Here we exploit the small fluctuations induced by the random local field driving the individual atoms $i$. As the local field has a random speckle structure (see Eq. (\ref{EqSpeckle})), the random light shifts and phases can be understood as an inhomogeneous broadening mechanism, depending on the interference term between the incident field $E_0(\mathbf r_i)$ and the scattered field $\varepsilon_i e^{i\varphi_i}E_0(\mathbf r_i)$.
The advantage of the light shift coupling and dephasing is the flexibility it offers as the laser intensity can be easily and quickly controlled. Fig. \ref{Fig4} shows the super- and subradiant decay rates as a function of the Rabi frequency $\Omega_0$ and the subradiant fraction (i.e. the remaining excited state population) $50 \, \Gamma^{-1}$ after switching off the laser (the laser was on during $50 \, \Gamma^{-1}$). When the Rabi frequency was varied we checked that the excited state population remains small ($\sum_i \rho_{i|i} < 0.15$) to ensure consistency with the model. We observed in Fig. \ref{Fig4} (b) up to a three time increase of the subradiant fraction when the intensity is raised compared to the $\Omega_0 \rightarrow 0$ value determined by Fano coupling.
Note that the subradiant fraction would be the same for any intensity with the effective Hamiltonian approach (due to linearity). Fig. \ref{Fig4} illustrates how changing laser intensity allows controlling the coupling strength between super- and subradiant modes and subsequently the subradiant population. As the laser is switched off, the local field, based on the interference between the incident and scattered field, is quickly reduced by a large amount. The inhomogeneous coupling is thus significantly decreased closing the `door' between the subradiant and superradiant states. One can see this effect in Fig. \ref{Fig4} (a), where a decrease in the superradiant decay rate is observed because there is still some inhomogeneous broadening source just after switching off the laser. However, for longer time the local field is much less intense and almost no variation of the subradiant decay rate is seen - light induced inhomogeneous broadening source is no longer present. In that case the subradiant decay rate is just the same as the one given by Fano coupling. In the same way, for low intensities, the inhomogeneous broadening induced by the laser is dominated by the Fano coupling. The subradiant decay and subradiant fraction then remain almost unaffected, as illustrated in Fig. \ref{Fig4} for  $\Omega_0 < 0.1 \, \Gamma$.

In conclusion, we have shown that inhomogeneous broadening schemes allow to understand and control storage of an optical excitation into `long-lived' subradiant modes.
We have proposed to use the cloud optical thickness, the driving laser intensity, or the cloud temperature as possible experimental control parameters for subradiance, but further parameters, such as a far detuned speckle field or magnetic field would yield similar results.
This opens the door for the first observation of Dicke subradiance of photons in a cold cloud of $N$ atoms in free space. Inhomogeneous broadening schemes will also be of interest to studies of Anderson localization of light in resonant two-level systems \cite{AGK2008}. Mapping the inhomogeneous coupling schemes described in this letter to a lambda scheme as used in quantum information science \cite{Cirac10} will help to understand limitations of storage of qubits in atomic vapors in modes less exposed to fast decay in a decoherence free subspace \cite{Ekert96}.
Controlled transfer to other modes can also be achieved with non-random coupling, as e.g. used in slow light experiments \cite{Weitz06}. Thus we expect that novel schemes to engineer more robust and faster storage and exploit the larger Hilbert space can be addressed using two-level systems as toy models.

\acknowledgments{We acknowledge fruitful discussions with the cold atom group at INLN, E. Akkermans and Ph. Courteille. Funding from IRSES project COSCALI and from USP/COFECUB is acknowledged.}

\end{document}